\documentclass[letterpaper,conference]{IEEEtran}
\usepackage{amsmath}
\usepackage{amsfonts}
\usepackage{graphicx}
\usepackage{bbold}

\begin{document}
\bstctlcite{IEEEexample:BSTcontrol}
\title{Similarity based hierarchical clustering of physiological parameters for
the identification of health states -- a feasibility study}

\author{\IEEEauthorblockN{Fabian Schrumpf\IEEEauthorrefmark{1}, Gerold
Bausch\IEEEauthorrefmark{1}, Matthias Sturm\IEEEauthorrefmark{1} and Mirco
Fuchs\IEEEauthorrefmark{1}} \IEEEauthorblockA{\IEEEauthorrefmark{1} Leipzig
University of Applied Sciences (HTWK), Laboratory for Biosignal Processing}}

\maketitle

\begin{abstract}
This paper introduces a new unsupervised method for the clustering of
physiological data into health states based on their similarity. We propose an
iterative hierarchical clustering approach that combines health states according
to a similarity constraint to new arbitrary health states. We applied our
method to experimental data in which the physical strain of subjects was systematically
varied. We derived health states based on parameters extracted from ECG data.
The occurrence of health states shows a high temporal correlation to the
experimental phases of the physical exercise. We compared our method to other clustering algorithms and found a significantly
higher accuracy with respect to the identification of health states.
\end{abstract} 
\section{Introduction}
Wearable sensor technologies are becoming more and more sophisticated. Nowadays
it is possible to conduct recordings of a multitude of physiological
signals (e.g. electrocardiogram, ECG) over several days by means
of extremely miniaturized battery-driven systems.
These systems are the basis for new clinical applications as, for instance,
vital parameter monitoring in patients long before and long after surgical
interventions. However, recording such data is useless without appropriate
processing and analyzing methods along with suitable visualization and alerting
strategies in case of relevant medical events. More precisely, it is necessary to (1) extract relevant physiological parameters, (2) to assess them not only according to some thresholds but also regarding their medical importance for the individual patient (that means with respect to the particular disease of the patient), and
(3) to derive some sort of classification of the current arbitrary health state
of a patient. Note that the health state should be understood as a certain
combination of vital parameters such as the heart rate (HR) and oxygen
saturation. It does not describe a particular condition of a patient (suffering from a certain disease or a certain degree of illness).

Recent advances in machine learning led to various applications of data mining
approaches for the prediction of adverse clinical events \cite{Kher2016}. These
approaches commonly rely on the definition and automated
adoption of thresholds.
They are based on the
extraction of vital paramters from physiological signals and their subsequent classification
according to predefined thresholds \cite{MohammadForkan2017}. However, it is
often difficult to define appropriate absolute thresholds, not least since labelled training
data is still limited and only available for certain diseases (e.g. specific
heart conditions) \cite{Goldberger2000}.
Hence, the general assessment of a patient’s health state during his/her daily
routine still remains an issue to be addressed.

It is very likely that similar health states can be described by a similar set
of physiological parameters. It might be useful to assess the
similarity between these parameter sets in order to identify similar health
states as well as changes from one health state to another.
We propose an iterative hierarchical clustering method which is based
on a similarity measure. If at least two parameter sets are found to be similar
according to this measure, they are assigned to the same health state. The
similarity can be assessed using various techniques. We utilized a rather
straightforward approach, i.e. the Mahalanobis distance. Our method is
unsupervised as it only relies on unlabelled data and does not need any
\textit{a priori} assumptions on the expected number of health states. It is
particularly suited for the analysis of data which arises, for instance, in a clinical setting.

Here, we present an experimental study to demonstrate the
ability of our method to identify correlates between a set of physiological
parameters and an arbitrary health state. Subjects were asked to perform
physical exercises in order to modulate their physical strain. We recorded ECG data and extracted several
parameters (e.g., respiratory rate). The automated clustering of health states
based on our approach revealed an excellent correspondence with the labelled
reference data, particularly in comparison to several other methods. The method
could thus be helpful in a clinical setting, where doctors are confronted with a
great variety of vital parameters gathered over long periods by providing alerts
to the staff when a multitude of vital parameters take unusual values. It could
be used to construct graphs that describe the alteration of health states over
time. From a medical perspective, this provides the opportunity to easily asses
the development oF a patients condition. It might also be possible to identify patterns of health states which inidcate a particular disease. The evalutation of our method in a clinical setting is part of our current research.

\section{Methods}
\subsection{Similarity based hierarchical clustering}
\begin{figure}[t] \centering
\includegraphics[width=0.9\columnwidth]{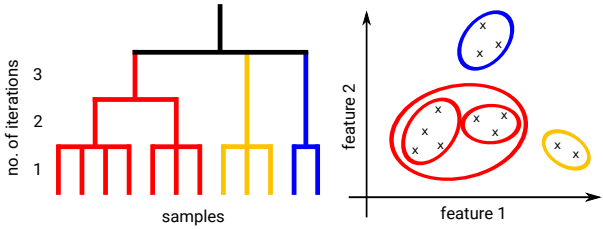} \caption{Left:
Example of the similarity based hierarchical clustering approach. A set of 12
samples is clustered into 3 health states. Right: The samples are combined based
on their Mahalanobis distance.}
\label{fig:tree_structure}
\end{figure}
Our method identifies biophysiological states by successively dividing a data
set into clusters based on their similarity. Given is the parameter \(x_n\) \(n
\in \mathbb{R}^M\), which contains \(M\) physiological parameters (i.e.
features) that are representative for sample \(n\) (i.e. time point) of the data
set consisting of N samples. We regard each of the samples in the data set as a
health state \(h_k\) \(k \in [1 \ldots N]\). The basic idea of our approach is
to combine different health states which are similar according to some
constraint to a new health state.
We expect that this leads to a small number of potential health states because
certain physical conditions can be characterized by a distinct set of parameter
values. More importantly, we believe that the variation of the similarity
constraint allows the identification of meaningful health states as, for
instance, sleep phases of a subject. We utilized the Mahalanobis distance as a
similarity measure since it allows to account for potential correlations between
the features. This is often the case in multinomial clustering problems where
features can differ in scale and variance among themselves
\cite{mahalanobis1936generalized}. Given two feature vectors \(x_i\) and \(x_k\)
and the covariance matrix of the underlying sample set \(C\) \(\in
\mathbb{R}^{N\times N}\), the Mahalanobis distance is defined as:
\begin{equation}
\label{eq:mahal_dist}
d(x_i,x_k) = \sqrt{(x_i - x_k)^T C^{-1}(x_i - x_k)}
\end{equation}
Note that, in the case of \(C = \mathbb{1}\), \(d(x_i,x_k)\) would be the
ordinary Euclidean distance between samples \(x_i\) and \(x_k\).

The first step
is the calculation of the pairwise Mahalanobis distance between all samples of the data set. Second, samples and their nearest neighbours for which the distance is lower than a certain threshold are assigned to the same
cluster (i.e., health state). Note that this means that not all samples in the
resulting cluster fulfill the pairwise similarity criterion. A small threshold
value represents a restrictive similarity constraint and prevents the generation
of large clusters in the first iteration step. The selection of this threshold is
essential for the ability to distingish particular states in the data set. Third, the centroids of all different clusters are computed. Each centroid serves as a representative for the
corresponding cluster and is used for subsequent iterations. Fourth, the
threshold (i.e., the similarity criterion) is relaxed by an amount of 5
percent in order to enable the clusters to grow. These steps are repeated until
an optimization criterion -- for instance a certain compactness of the clusters
-- is reached. In our feasibility study, however, we did not use such an
optimization criterion. For the sake of simplicity, we repeated this procedure
until a predefined number of iterations was reached.

The proposed algorithm combines the data into a tree-like structure and is
therefore related to classical agglomerative clustering. The main difference is
that the resulting tree is not binary. This can be seen in Fig.
\ref{fig:tree_structure}. 
\subsection{Experimental study}
\label{sec:experimental_study}
\begin{figure}[t] \centering
\includegraphics[width=0.9\columnwidth]{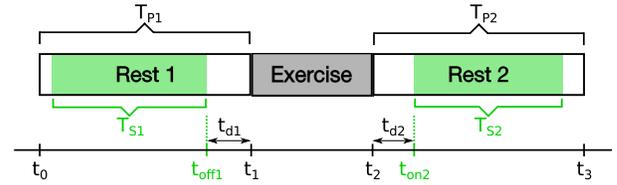}
\caption{The
experimental procedure consists of two resting phases (5 minutes each)) and a
phase of physical strain (3 minutes). The end of each phase is
indicated by means of a trigger button (\(t_0\), \(t_1\), \(t_2\), \(t_3\)).
The following measures where derived: The duration of each detected resting
state (\(T_{S1}\), \(T_{S2}\)), the duration of each true resting phase
(\(T_{P1}\), \(T_{P2}\)). Offset of the first resting state (\(t_{off1}\)),
onset of the second resting state \(t_{on2}\).}
\label{fig:experimental_protocol}
\end{figure}
\begin{figure*}[ht]
\centering
\includegraphics[width=\textwidth]{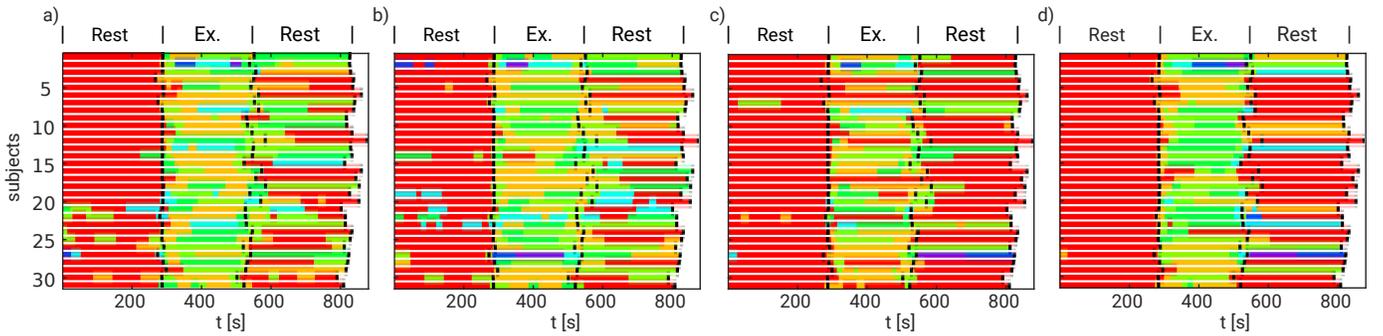}
\caption{Clustering result by means of the ECG-based features (respiration
frequency, respiration depth, HRV, Q-TE duration) using a) Euclidean distance
based fuzzy clustering (FCeucl); b) Mahalanobis distance based fuzzy clustering
(FCmahal); c) binary hierarchical clustering (BiTree)); d) similarity based
hierarchical clustering (Htree). Each line in the plots corresponds to the
clustering results of a single subject.}
\label{fig:health_states_raw}
\end{figure*} 
We carried out an experimental study in order to examine the feasibility of our
approach. The aim was to identify different phases of physical strain by means
of the proposed clustering approach. We recruited 32 healthy subjects among the
employees and students of our research group. Their mean age was 26.8 years (SD
= 5.5 years). The experiment consists of three different phases which are
illustrated in Fig. \ref{fig:experimental_protocol}. The subjects remained in
a supine position and where told to relax during the resting phases. They were
asked to perform squats in order to raise their respiratory frequency and heart
rate during the exercise phase. Single channel ECG data was recorded using a PLUX biosignals
researcher measurement system (PLUX wireless biosignals, Lisbon, Portugal).
The phases were labelled by means of a trigger button.

The ECG recordings were used to extract features to
distinguish between different states of physical strain. For instance, increased
respiratory rates and shortened Q-TE durations are known to be correlated to physical strain \cite{Hijzen1985}. In a preprocessing
step, baseline removal was applied and the 50 Hz powerline interference was
removed  \cite{Zhang2014a}.
The signal was then subdivided into windows of 30 s length and an overlap of 29
s between adjacent windows. The following features were extracted in each
window: (i) respiration rate, (ii) respiration depth, (iii) heart rate
variability (HRV) and (iv) timing lag between the onset of the QRS-wave and
the offset of the T-wave (Q-TE) \cite{Hijzen1985}. In order to estimate
the HRV, the QRS complexes including their on- and offsets were
extracted \cite{Li1995,Manriquez2007}. The respiratory signal was estimated by
interpolating the amplitudes of the R-peak \cite{Moody1985}. The depth of respiration
was then calculated as the root mean square value of the respiratory signal. The
respiratory rate was extracted by means of the autocorrelation function of the
respiratory signal \cite{Schrumpf2016a}. The T-wave detection was performed according to the
algorithm described in \cite{Chatterjee2012}.

These four features were the basis for the subsequent hierarchical clustering of
each individual data set. We used a total of 35 iterations. The threshold was
set to \(d_\mathit{thresh} = 0.05\,max(d_\mathit{mahal})\) for the first iteration.
This value was increased by 5\% before each iteration.

We compared our method to three other clustering algorithms. First, a fuzzy clustering method based on the
Gustafson-Kessel-Algorithm (FCeucl) was used \cite{Gustafson1978}. This is a
top down approach which optimizes the probability distribution of each sample to belong
to a certain cluster. The algorithm is based on the \textit{a priori} definition
of the number of expected clusters. The probability estimation is based on
the Euclidean distance between the sample and the centres of the clusters. The
second method is a variation of the first one. It uses Mahalanobis distances
(FCmahal) rather than Euclidean distances \cite{Zhao2015}. The third method is a
classical hierarchical clustering approach that groups samples over a variety of
distances by building a binary tree (BiTree).

Finally, the resulting clusters and their occurrences during the experiment were
evaluated to extract the duration of the three different phases from the
individual data sets. We expected to find two major health states that coincide
with the physical strain during the three experimental phases. First, a resting state during the first relaxation phase, a stress state during
physical exercise, and another resting state in the second relaxation phase.
We identified the dominant cluster in each phase and computed its
start and end time. We compared these times with those obtained on the
basis of the trigger button. We calculated the time difference \(t_{d1}\) (Eq.
\ref{eq:td1}) between the end of the
dominant cluster and the time of the first button press at time \(t_1\) (see
also figure \ref{fig:experimental_protocol}).
This indicates how accurately the algorithm detects changes in physical strain. For the same
reason, the duration \(t_{d2}\) (Eq. \ref{eq:td2}) between the beginning  of the
second resting state and the second button press at time \(t_2\) was determined.
Additionally, the duration of the resting phases based on the clustering approaches and based on the trigger
channel were compared (\(_\Delta T_1\) and \(_\Delta T_2\) see Eq. \ref{eq:dT1} and
\ref{eq:dT2}).

\begin{minipage}{0.45\columnwidth}
\begin{equation}
t_{d1} = t_1 - t_{\mathit{off1}}
\label{eq:td1}
\end{equation}
\begin{equation}
t_{d2} = t_{\mathit{on2}} - t_2
\label{eq:td2}
\end{equation}
\end{minipage}
\begin{minipage}{0.5\columnwidth}
\begin{equation}
_\Delta T_1 = T_\mathit{P1} - T_{S1}
\label{eq:dT1}
\end{equation}
\begin{equation}
_\Delta T_2 = T_\mathit{P2} - T_{S2}
\label{eq:dT2}
\end{equation}
\end{minipage}
\subsection{Sensitivity Analysis}
We performed a sensitivity
analysis in order to evaluate the influence of the threshold parameter
\(d_\mathit{thresh}\) on the clustering results obtained with the HTree method.
The threshold parameter was varied in a range from \(d_\mathit{thresh} = 0.1 \ldots 0.8\). The times and
durations of the major phases extracted using these different thresholds were compared to results
obtained by the setup (\(d_\mathit{thresh} = 0.5\)) described in section
\ref{sec:experimental_study}. This particular threshold value was empirically determined and leads to the best clustering results of the data sets.
\section{Results}
\begin{figure}[b]
\centering
\includegraphics[width=0.9\columnwidth]{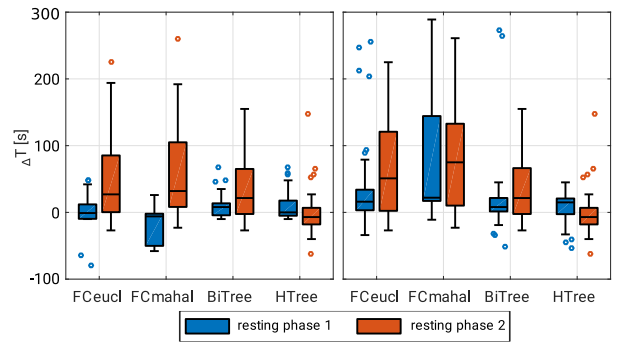}
\caption{Left: Delay between resting state offset and first button press during
resting phase 1 (\(t_{d1}\) blue) and third button press and resting state onset
during resting phase 2 (\(t_{d2}\) orange); Right: duration difference between
first resting state and resting phase 1 (\(_\Delta T_1\) blue) and between
second resting phase and resting phase 2 (\(_\Delta T_2\) orange)}
\label{fig:health_states_boxplots}
\end{figure}
\begin{figure*}[ht]
\centering
\includegraphics[width=\textwidth]{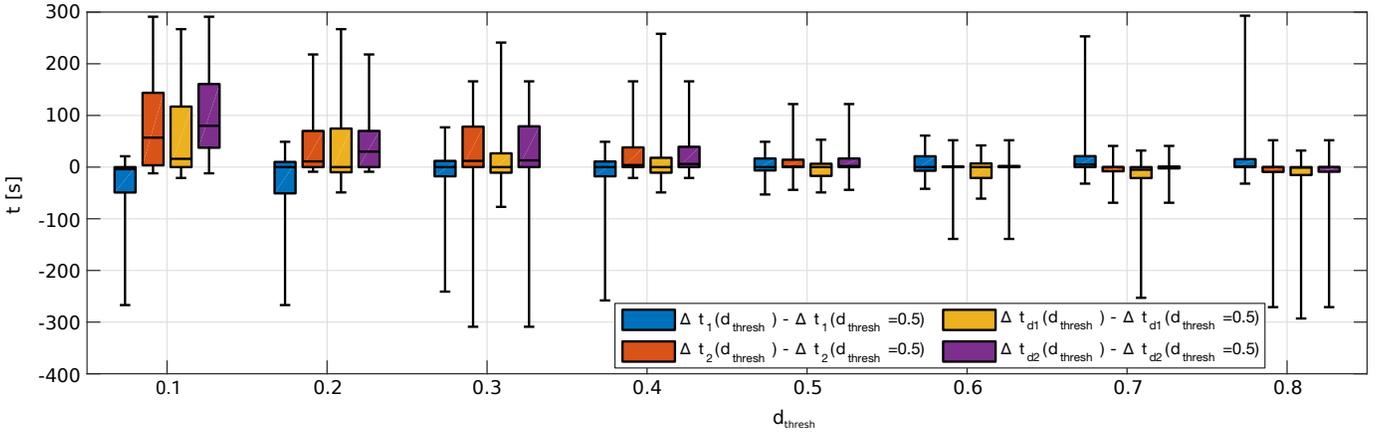}
\caption{Results of the sensitivity analysis in order to investigate the
influence of \(d_\mathit{thresh}\) on the clustering results. The threshold
parameter was varied in a range of \(t_\mathit{thresh} = 0.1 \ldots 0.8\) and
the delay times(Eq. \ref{eq:td1} and \ref{eq:td2}) as well as the phase durations (Eq.
\ref{eq:dT1} and \ref{eq:dT2}) were compared to the ones obtained with
\(d_\mathit{thresh}=0.5\).}
\label{fig:sens_analysis}
\end{figure*}
Fig. \ref{fig:health_states_raw} shows the identified clusters (color-coded) over time for all
methods and subjects. Each black marker represents a button press and indicates
the true end of a particular experimental phase (cmp. Fig. \ref{fig:experimental_protocol}). It can
be seen that all methods establish an initial state (denoted in red) during the first resting
phase almost consistently for all subjects. This can be explained with the fact that the
subjects lie in a supine position and are relaxed which leads to a low
variation of the underlying physiological parameters. During the exercise phase
the human body reacts to the increased physical strain with an increased
respiration frequency and respiration depth, a stronger HRV and shorter
Q-TE-durations. It can be seen that the algorithms consistently detect a change
of the health state in a high temporal correlation to the first button press.
That means that the beginning of the exercise phase is detected correctly.
However, the states are much less stable and even additional states are
introduced. After the exercise phase is completed (second black marker) the
subjects relax during the second resting phase. We expected that the resulting health state resembles the state from the first
resting phase. The results for the clustering methods BiTree and
HTree clearly confirm our expectations for most subjects. However, there are
some subjects (e.g. subject 3) for which the health state in the second resting phase
is not equivalent to the state in the first resting phase. Instead, an
additional state is introduced which seems to be in temporal
correlation with the duration of the second resting phase. This effect occurs
even more frequent in the clustering methods FCeucl and FCmahal.

Next, a quantitative evaluation of the results was carried out. We analysed the
temporal consistency between the clustering results and the experimental procedure. In particular, we compared the time of the end of the
first resting state obtained by means of the button press (\(t_1\)) and on the
basis of the clustering approaches (see Fig. \ref{fig:health_states_boxplots}
left).
Likewise, we compared the times of the beginning of the second resting state
(see Fig. \ref{fig:health_states_boxplots} left). Furthermore we
calculated the duration of the resting states obtained using the clustering
methods and compared them as well (Fig.
\ref{fig:health_states_boxplots}, right). These values serve as a measure of how accurately the methods follow the changes in physical activity.

The results indicate that most of the algorithms detect a health state change in
temporal proximity to a true change of the physical activity. It can be seen in
Fig. \ref{fig:health_states_boxplots} (left panel) that the time differences
obtained for method HTree are particularly small. In other words, the method provides a relatively good detection of the
beginning and the end of the exercise phase. Likewise, the methods FCeucl and
BiTree provide a good detection of the beginning of the exercise phase but
clearly perform weaker regarding the correct detection of the beginning of the second
resting phase. The time differences obtained for method FCmahal are large
compared to all other methods.

The results obtained from analyzing the duration of the resting phases are
depicted in Fig. \ref{fig:health_states_boxplots} (right panel). It can be seen
that the durations obtained for algorithm HTree are very similar to the true duration
of the experimental phases.
The methods FCeucl and BiTree provide a good estimation for the duration of the
first resting phase. However, the difference between the true and the estimated
duration of the second resting phase is rather large for these two methods.
Likewise to the previous analysis, method FCmahal provides the weakest
performance and leads to large time differences.

We performed two-sample Kolmogorov-Smirnov tests to check whether the
results obtained by the HTree method are significantly different from the ones
obtained from the other methods. The results can be seen in table
\ref{tab:p-values}. It can be concluded that the HTree method performs
significantly better regarding the detection of the onset and
duration of the second resting state compared to other approaches.
\begin{table}[ht]
\caption{p-values of the two-sample KS-Test checking whether the times obtained
with the BiTree method are significantly different from the times obtained by
the other methods}
\label{tab:p-values}
\centering
\begin{tabular}{c c c c}
\hline
& FCeucl & FCmahal & HTree\\
\hline
\(t_{d1}\) & 0,36 & \textit{\(<0.05\)} & 0.14\\
\(t_{d2}\) &  \(<0.05\) & \(<0.05\) & \(<0.05\)\\
\(_\Delta T_1\) & 0,21 & \(<0.05\) & 0,15 \\
\(_\Delta T_2\) & \(<0.05\) & \(<0.05\) & \(<0.05\)\\
\hline
\end{tabular}
\end{table}

We conducted a sensitivity analysis to investigate how a particular
threshold parameter \(d_\mathit{thresh}\) influences the clustering
results. The parameter was systematically varied and the clustering results
(eg.
\(t_{d1}\), \(t_{d2}\), \(_\Delta T_1\) and \(_\Delta T_2\)) were compared to
the results obtained for \(d_\mathit{thresh} = 0.5\). Figure
\ref{fig:sens_analysis} shows that a small value leads to huge 
deviations from the optimum \(d_\mathit{thresh} = 0.5\). A larger threshold
value would be less problematic.
\section{Discussion}
We have introduced a new method  for the identification of discrete individual
health states. Our approach is based on determining the similarity between a
certain set of physiological parameters at different time points. This leads to
an iterative hierarchical clustering procedure in which existing health states
are combined into new health states. We proposed to use
the Mahalanobis distance to estimate similarities but different measures
might be applied as well.
We have proven the general feasibility of our approach by applying the method to
experimental data. We conducted an experimental study in which subjects
underwent different levels of physical strain. The clustering of health
states was performed using our proposed approach and three reference methods.

The results obtained with our proposed procedure revealed a high temporal
correlation between the derived health states and the true
experimental phases. More importantly, we identified phases with stable
health states before and after the exercise phase consistently over subjects.
These two states were identical in most of the subjects. There are different
reasons why this was not the case in some subjects. First, we did not account
for the particular physical condition when recruiting the subjects. Some subjects might need a particularly long recovery time which can lead to different health
states after exercise. Likewise, the absence of a resting
phase prior to the experiment might cause a higher initial level of physical
stress at the beginning of the experiment and, therefore, a different health
state.

We have shown that our method shows a
significantly better performance compared to the reference methods, particularly regarding  the \(_\Delta T_2 \) and the
\(t_{d2}\) measure.
The measures \(_\Delta T_1\) and \(t_{d1}\), though, only provide a
significantly better performance when compared to the FCmahal method. We
conclude that all considered methods are able to detect a transition from a resting state to an exercise
state more or less precisely. However, only the method proposed here leads to a
similar classification of the health states before and after the experimental
phase consistently across nearly all subjects. It should be noted that our
experimental design is simple in a sense that it allowed us to distinguish
between two very different states of physical strain very clearly. It is not in
the scope of the present work to discriminate between different levels of
physical exercise. 

It is straightforward to extend our method in several ways. An important issue
is the selection of a criterion to stop the iterative procedure. The utilization
of a more specific criterion (e.g. cluster compactness or a certain cost
function that is being optimized) might lead to better clustering results.
Another important issue is the selection of the features as a basis for
clustering.\newpage
It is obvious that the parameters that were selected here are sufficiently suited to explain the subjects health states during the conducted
experiment. However, other applications might need additional
information. This can be drawn, for example, from alternative physiological
parameters or even from non-physiological features as, for instance, behavioural
data. Another aspect worth investigating is the performance of our method in a
typical machine learning setting where a defined subset of subjects serves as a
training set and the learned cluster parameters are applied to a different set
of subjects. This would also require additional information on the subject
(e.g. body mass, stature). 

Our final conclusion is that our method is applicable for the identification of
health states based on physiological data. Apart from the detection of physical
strain, there are other relevant application scenarios as, for
example, the detection of pain from physiological and behavioural data.
\section*{Acknowledgments}
This work was funded by the German Federal Ministry of Education and Research
(BMBF) (FKZ 03FH032IX5).
\bibliographystyle{IEEEtran}
\bibliography{IEEEabrv,literature}
\end{document}